\newcommand\mM{\ifmmode(m{-}M)\else$(m{-}M)$\fi}

\newcommand\zacs{\ifmmode z_{850}\else$z_{850}$\fi}
\newcommand\iacs{\ifmmode i_{775}\else$i_{775}$\fi}
\newcommand\gacs{\ifmmode g_{475}\else$g_{475}$\fi}
\newcommand\racs{\ifmmode r_{625}\else$r_{625}$\fi}
\newcommand\vacs{\ifmmode V_{606}\else$V_{606}$\fi}
\newcommand\gz{{\ifmmode{(g_{475}{-}z_{850})}\else$(g_{475}{-}z_{850})$\fi}}
\newcommand\gzacs{{\ifmmode{g_{475}{-}z_{850}}\else$g_{475}{-}z_{850}$\fi}}
\newcommand\riacs{{\ifmmode{r_{625}{-}i_{775}}\else$r_{625}{-}i_{775}$\fi}}
\newcommand\rzacs{{\ifmmode{r_{625}{-}z_{850}}\else$r_{625}{-}z_{850}$\fi}}
\newcommand\izacs{{\ifmmode{i_{775}{-}z_{850}}\else$i_{775}{-}z_{850}$\fi}}
\newcommand\vzacs{{\ifmmode{V_{606}{-}z_{850}}\else$V_{606}{-}z_{850}$\fi}}
\newcommand\vi{{\ifmmode{(V{-}I)}\else$(V{-}I)$\fi}}

\newcommand\mbari{\ifmmode\overline{m}_I\else$\overline{m}_I$\fi}
\newcommand\mbarI{\ifmmode\overline{m}_I\else$\overline{m}_I$\fi}
\newcommand\mbarz{\ifmmode\overline{m}_z\else$\overline{m}_z$\fi}
\newcommand\mbar{\ifmmode\overline{m}\else$\overline{m}$\fi}

\newcommand\Mbar{\ifmmode\overline{M}\else$\overline{M}$\fi}
\newcommand\lbar{\ifmmode\overline{L}\else$\overline{L}$\fi}
\newcommand\Mbarz{\ifmmode\overline{M_z}\else$\overline{M}_z$\fi}

\documentclass[
    ,final            
  ,numberedheadings 
  ]
  {aipproc}

\layoutstyle{6x9}

\begin{document}

\title{The Mass-Metallicity relation
in galaxies of different
morphological types}


\author{F. Calura}{
  address={Dipartimento di Astronomia, Universit\'a di Trieste, 
		Via G.B. Tiepolo 11, 34143 Trieste, Italy},
  email={fcalura@oats.inaf.it}
}

\iftrue
\author{A. Pipino}{
  address={Department of Physics \& Astronomy , University of Southern California, Los Angeles 90089-0484, USA}}

\author{F. Matteucci}{
address={Dipartimento di Astronomia, Universit\'a di Trieste, 
		Via G.B. Tiepolo 11, 34143 Trieste, Italy}}

\author{C. Chiappini}{
  address={Observatoire de Gen\`eve, Universit\`e de Gen\`eve, 51 Chemin des Maillettes, CH-1290 Sauverny, Switzerland},
altaddress={INAF- Osservatorio Astronomico di Trieste, Via G. B. Tiepolo 11, 34131 Trieste, Italy}}

\author{R. Maiolino}{
  address={INAF - Osservatorio Astronomico di Roma, via di Frascati 33, 00040 Monte Porzio Catone, Italy}}

\author{N. Menci}{
  address={INAF - Osservatorio Astronomico di Roma, via di Frascati 33, 00040 Monte Porzio Catone, Italy}}

\fi


\begin{abstract}
By means of chemical evolution models of different morphological types, 
we study the mass-metallicity (MZ) relation and its evolution with redshift. 
Our aim is to  
understand the role of galaxies of different morphological types in 
the MZ relation  at various redshift. One major result is that at high redshift, 
the majority of the galaxies falling on the MZ plot
are apparently proto-ellipticals. 
Finally, we show some preliminary results of a study of the MZ relation in a framework of 
hierarchical galaxy formation. 
\end{abstract}

\date{\today}

\maketitle

\subsection{The redshift evolution of the MZ relation and the morphology of star-forming galaxies}
\label{monolithical}
Various theoretical interpretations have been proposed so far to explain the MZ relation.  
The MZ relation can be reproduced by means of (i) particularly efficient 
outflows of metal-enriched gas in dwarf galaxies (Larson 1974; Erb et al. 2006); 
(ii) star formation efficiencies increasing with galactic mass (a.k.a. downsizing, Lequeux et al. 1979, Matteucci 1994); 
(iii) by adopting  a larger upper mass cutoff in the initial mass function (IMF) of larger galaxies (K\"oppen et al. 2007). \\
In this paper, we use chemical evolution models to  
understand the role of galaxies of different morphological types in the MZ relation  at various redshift. 
The chemical evolution models for ellipticals and spirals are designed to reproduce  all 
their main chemical properties, as well as other phyisical parameters such as gas fractions and 
present-day Supernova (SN) rates  
(Chiappini et al. 2003; Calura \& Matteucci 2004;  Pipino \& Matteucci 2004; Calura \& Matteucci 2006, Cescutti et al. 2007). 
For each morphological type, we start from three baseline models of three different pesent-day stellar masses. 
We choose a redshift of formation $z_f$ and an age dispersion $\Delta_t$. We simulate  
a continuous galaxy formation process 
across a given time interval computed as a function of $t(z_f)$, the age of the universe at the redshift $z_f$, 
and $\Delta_t$. Within this time interval, 
we extract randomly ages and we compute
the relevant physical quantities at these ages. In this way, 
we can simulate a continuous galaxy formation process and we can reproduce a fine mass grid. \\
%
%

\begin{figure}
  \includegraphics[height=6cm,width=8cm]{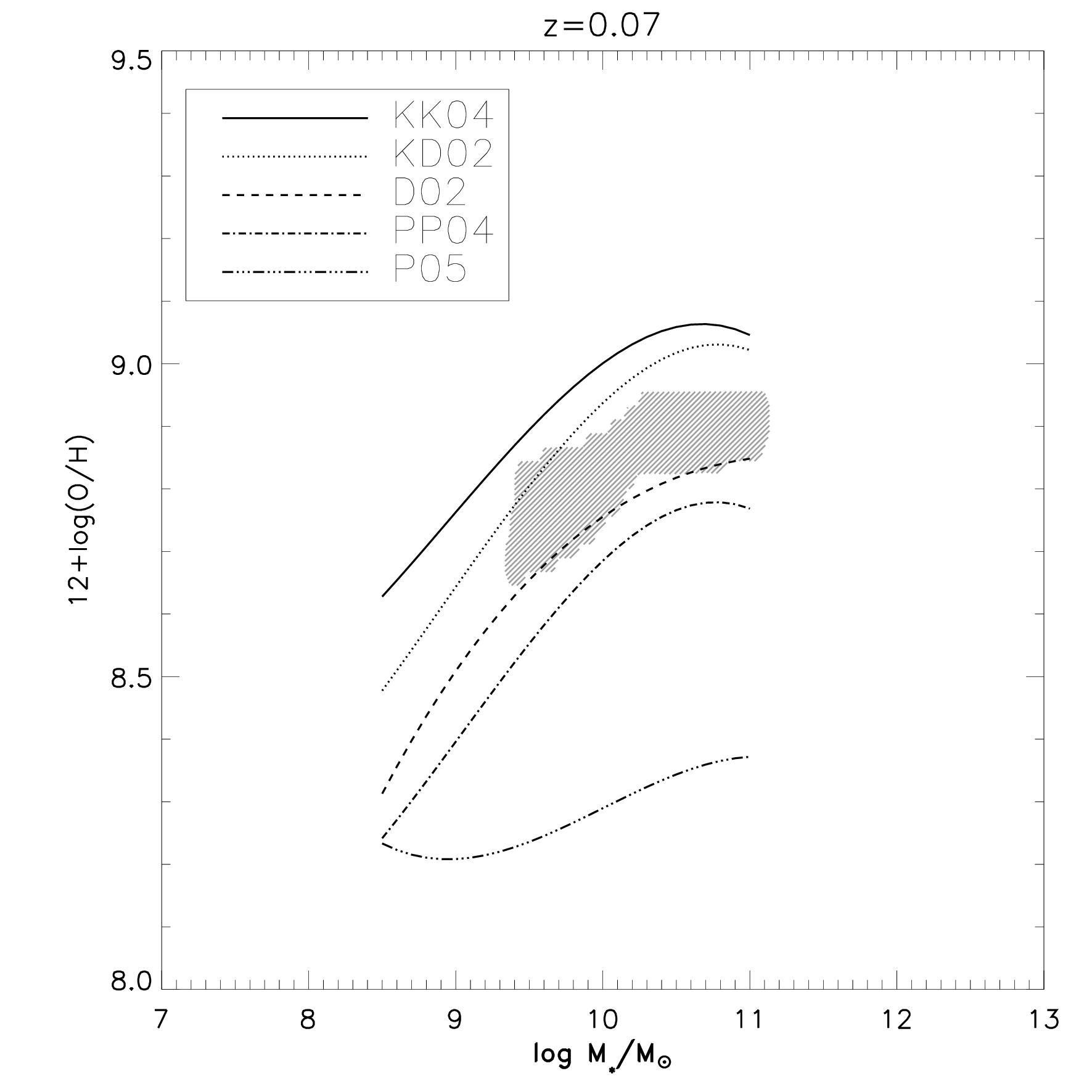}
  \includegraphics[height=6cm,width=8cm]{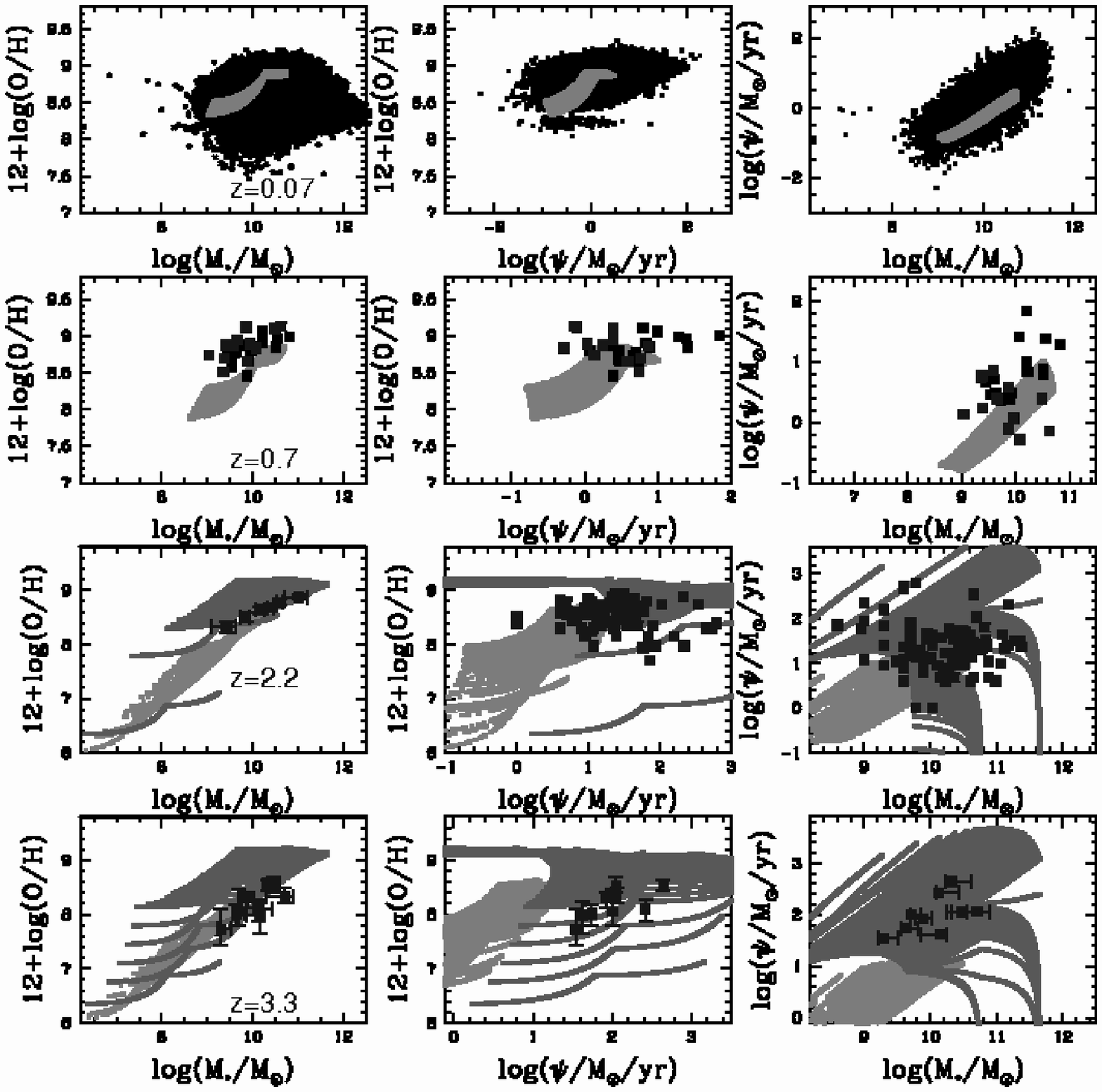}
  \caption{Left panel: Predicted Mass-Metallicity for spiral galaxies (shaded region) at $z=0.07$. 
The lines of different types are best-fits to the observational MZ relation as 
calculated by Kewley \& Ellison (2008) using different 
metallicity calibrations. KK04: Kobulnicky \& Kewley (2004); KD02:  Kewley \& Dopita (2002); D02: Denicol\'o (2002); 
PP04: Pettini \& Pagel (2004); P05: Pilyugin et al. (2005). Right panel: redshift evolution of the MZ, 
(O/H) vs SFR and SFR vs mass plots for ellipticals and spirals  
and as observed by various authors at $z=0.07$, $z=0.7$, $z=2.2$, $z=3.3$ (se text for further details). }
\label{fig1}
\end{figure}

In this paper, we are interested in the MZ relation of only star-forming galaxies. 
In Fig.~\ref{fig1}, left panel, 
we show the predicted MZ relation for local spirals (shaded region), computed assuming $z_f=3$ and $\Delta_t=5$ Gyr, 
compared with the fits to the 
observational MZ relations at $z=0.07$ published by Kewley \& Ellison (2008), obtained assuming different metallicity calibrations, 
indicated by the various lines. 
We show the predictions only for spirals 
since in local ellipticals star formation is negligible. 
The parameters $z_f=3$ and $\Delta_t=5$ 
have little effect on the zero point and on the slope of the predicted 
mass-metallicity relation. 
We can see that it is possible to reproduce the 
MZ relation in spirals by means of an increasing star formation efficiency as a function of the stellar mass. 
The use of various  calibrations 
produces very different results, concerning both 
the zero-point and the slope of the observational MZ relation.
Our predictions are compatible with results of the KD02 and D02 calibrations. \\
In Fig.~\ref{fig1}, right panel, we show the redshift evolution of the MZ relation 
(first column on the left), the O/H vs SFR relation (second column) and 
of the SFR vs $M_{*}$ (third column). The four rows of the figure refer to different 
redshifts at which stellar masses, insterstellar metallicities and star formation 
rates have been measured for various galaxy samples. 
The black squares are the observations at various redshifts (see Maiolino et al. 2008). 
The dark grey regions are the predictions for ellipticals, whereas the light-grey 
areas are the predictions for spiral galaxies. 
The predictions are computed assuming $z_f=3$ and $\Delta_t=5$ Gyr for spirals and 
$z_f=3$ and $\Delta_t=3$ Gyr for ellipticals.  
Such  values of $\Delta_t$ allow us to produce present-day spirals and ellipticals with ages 
compatible with observational estimates (e. g., Bernardi et al. 1998).  
At $z=3.3$, the majority of the observed galaxies are apparently proto-ellipticals. 
At $z=2.2$, a morphological mix of spirals and ellipticals can reproduce the observational data. 
At $z=0.7$, according to our predictions, only spirals are star-forming. 
We slightly underestimate the metallicities in the MZ plot, 
however, given the uncertainties due to calibration discussed above,we do not consider this a major issue. 
At this redshift, we also underestimate most of the observed star formation rates. 
This discrepancy may be due to episodic starbursts, possibly triggered by a dynamical process such as galaxy interactions, 
which in our models are not taken into account. Similar conclusions can be drawn from the three plots at $z=0.07$. 
 
\subsection{The MZ relation in hierachical framework}
\begin{figure}
  \includegraphics[height=6cm,width=8cm]{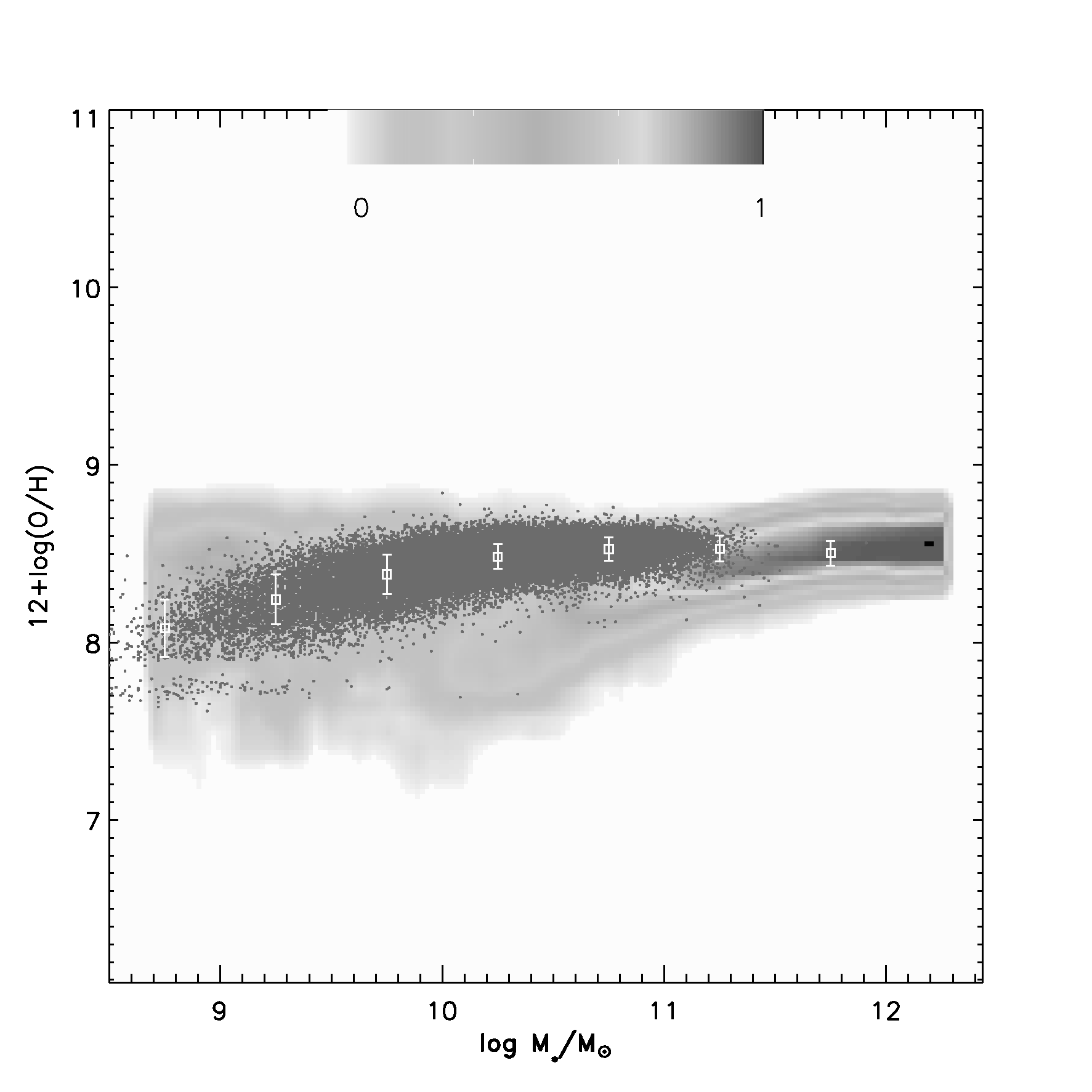}
  \includegraphics[height=6cm,width=8cm]{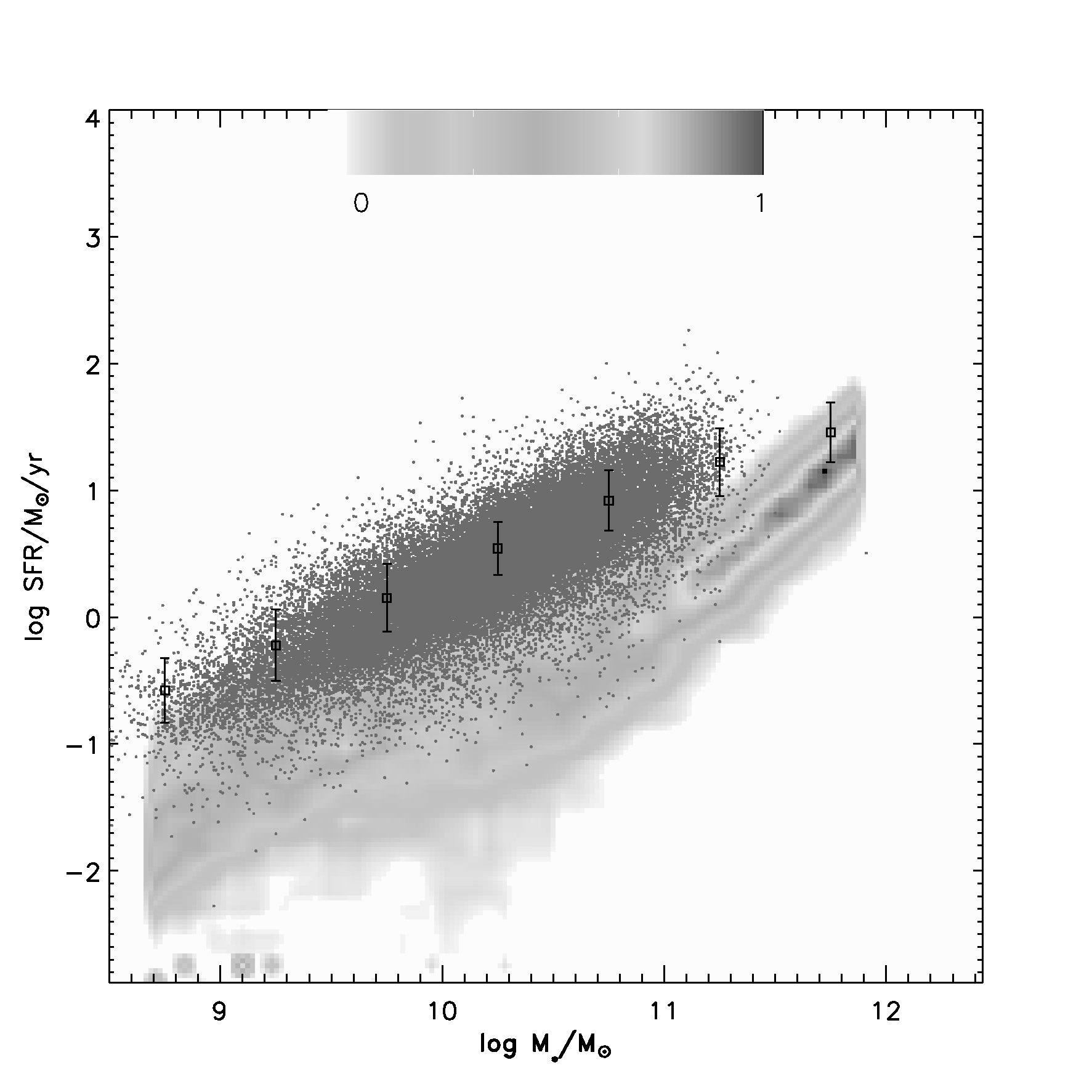}
  \caption{Left panel: Predicted MZ relation at $z=0.07$ by means of the SAM model of Menci et al. (2002). 
The predicted density of galaxies scales with the grayscale bar on the top. The observational data (grey points) 
are from Kewley \& Ellison (2008). The white points with the error bars are the average values of the observations 
with their $1-\sigma$ in various mass bins. Right panel: Predicted and observed 
SFR vs Mass  relation at $z=0.07$. Predictions and observational data as in the left panel. }
\label{fig2}
\end{figure}

In this section, we describe some results concerning the local MZ relation obtained by means of a hierarchical 
semi-analytical model (SAM) for galaxy formation. 
This model is described in Menci et al. (2002) and Calura et al. (2004). 
Some of its main features are the inclusion of a physical model for AGN feedback (Menci et al. 2008) and 
a physical description of starbursts triggered by galaxy interactions (Menci et al. 2004). 
In this case, it is not possible to infer the morphology of the galaxies falling 
on the MZ relation. However, a study of the MZ relation performed by means of an ab-initio galaxy formation model may 
be useful to derive contraints for some of the parameters involved in galaxy formation studies. 
In the left panel of Fig.~\ref{fig2}, 
the predicted density of galaxies scales with the grayscale bar on the top. The observations are instead indicated by 
the grey points. The observed metallicities have been shifted downwards in order to match our predicted zero point 
of the MZ relation. 
We can see that the dispersion of the observational data is overestimated by our predictions. The reason is that 
we overestimate the number of metal-rich local dwarf galaxies. This is a well-known problem of any  
 galaxy formation model based 
on a $\Lambda-$CDM cosmology.  The metallicity and the gas content of dwarf galaxies depend strongly on the feedback parameter: in principle, 
by varying this parameter, it may be possible to reduce the metallicity of dwarf galaxies. This aspect will be investigated in the future.
In the right panel of Fig.~\ref{fig2}, we show the predicted SFR vs mass at $z \sim 0.1$, 
compared with the local SDSS measures. 
Also in this case the SFRs are underestimated by our predictions in the stellar mass range 
covered by the observations. 
This shows that it is possible that the SDSS 
SFRs considered here may represent overestimates, possibly 
owing to overestimated dust attenuation in the SDSS sample (J. Brinchmann, provate communication). This 
aspect will be fixed with the next SDSS data release. 
In the future, we plan to  refine 
our study of the MZ relation by means of the semi-analytical galaxy formation model, assessing the roles of fundamental  parameters such as the feedback and the IMF. 

\begin{theacknowledgments}
F.C. acknowledges financial support from contract ASI-INAF I/016/07/0.
F.C. and F. M. acknowledge financial support from PRIN2007, Prot.2007JJC53X\_001.
C.C. acknowledgs financial support from Swiss National Science Foundation.
\end{theacknowledgments}



\begin{thebibliography}{9}
\bibitem{1} Bernardi  M., Renzini  A., da Costa  L. N., Wegner  G., Alonso  M. V., Pellegrini  P. S., Rité  C., Willmer  C. N. A., 1998, ApJ, 508, L43 
\bibitem{2} Bower R. G.; Lucey J. R.; Ellis R. S., 1992, MNRAS, 254, 601
\bibitem{3} Calura  F., Matteucci  F., 2004, MNRAS, 350, 351 
\bibitem{4} Calura F.,  Matteucci F., 2006, ApJ, 652, 889
\bibitem{5} Calura  F., Matteucci  F., Menci  N., 2004, MNRAS, 353, 500    
\bibitem{6} Cescutti G.; Matteucci F.; François P.; Chiappini C.,  2007, A\&A, 462, 943
\bibitem{7} Chiappini  C., Romano  D., Matteucci  F., 2003, MNRAS, 339, 63 
\bibitem{8} Denicol\'o G.; Terlevich R.; Terlevich, E., 2002, MNRAS, 330, 69
\bibitem{9} Erb D. K., et al., 2006, ApJ ,644, 813
\bibitem{10} Kewley L. J.; Dopita M. A., 2002, ApJS, 142, 35
\bibitem{11} Kewley L. J.; Ellison S. L., 2008, ApJ, 681, 1183
\bibitem{12} Kobulnicky H. A.; Kewley L. J., 2004, ApJ, 617, 240
\bibitem{13} K\"oppen, J., Weidner  C., Kroupa P., 2007, MNRAS, 375, 673 
\bibitem{14} Larson  R. B., 1974, MNRAS, 169, 229
\bibitem{15} Lequeux J., Peimbert M., Rayo J. F., Serrano A., Torres-Peimbert S., 1979, A\&A, 80, 155 
\bibitem{16} Matteucci F. 1994, A\&A, 288, 57
\bibitem{17} Menci N.; Cavaliere A.; Fontana A.; Giallongo E.; Poli F., 2002, ApJ, 575, 18
\bibitem{18} Menci N.; Cavaliere A.; Fontana A.; Giallongo E.; Poli F.; Vittorini V., 2004, ApJ, 604, 12 
\bibitem{19} Menci N.; Fiore F.; Puccetti S.; Cavaliere A., 2008, ApJ, 686, 219
\bibitem{20} Maiolino R., et al., 2008, A\&A, 488, 463
\bibitem{21} Pettini M., Pagel B. E. J., 2004, MNRAS, 348, L59 
\bibitem{22} Pilyugin L. S.; Thuan T. X., 2005, ApJ, 631, 231
\bibitem{23} Pipino  A., Matteucci  F., 2004, MNRAS, 347, 968 
\end{thebibliography}

\end{document}